\documentclass[aj,twocolumn]{aastex62}

\usepackage{placeins}
\usepackage{comment}

\newcommand{\kms}{\ensuremath{\rm km\,s^{-1}}}
\newcommand{\ms}{\ensuremath{\rm m\,s^{-1}}}

\newcommand{\postlambda}{1_{-33}^{+41}}
\newcommand{\postthreed}{2_{-23}^{+27}}

\newcommand{\teff}{\ensuremath{T_{\rm eff}}}

\newcommand{\vsini}{\ensuremath{v \sin{I_\star}}}

\newcommand\mysim{\mathord{\sim}}

\newcommand{\rstar}{\ensuremath{R_\star}}

\newcommand{\rearth}{\ensuremath{R_\earth}}
\newcommand{\mearth}{\ensuremath{M_\earth}}

\newcommand{\rpl}{\ensuremath{R_{p}}}

\received{\today}
\submitjournal{ApJL}

\shorttitle{Prograde orbit for TOI-942b}
\shortauthors{Wirth et al.}

\begin{document}

\title{TOI-942b: A Prograde Neptune in a $\mysim60$ Myr old Multi-transiting System\footnote{This paper includes data gathered with the 6.5 meter Magellan Telescopes located at Las Campanas Observatory, Chile.}}

\correspondingauthor{Christopher P. Wirth}
\email{cwirth@college.harvard.edu}

\author[0000-0003-1656-011X]{Christopher P. Wirth}
\affiliation{Harvard University, Cambridge, MA 02138, USA.}
\affiliation{Center for Astrophysics \textbar{} Harvard \& Smithsonian, 60 Garden St., Cambridge, MA 02138, USA.}

\author[0000-0002-4891-3517]{George Zhou}
\affiliation{University of Southern Queensland, Centre for Astrophysics, West Street, Toowoomba, QLD 4350 Australia}
\affiliation{Center for Astrophysics \textbar{} Harvard \& Smithsonian, 60 Garden St., Cambridge, MA 02138, USA.}

\author[0000-0002-8964-8377]{Samuel N. Quinn}
\affiliation{Center for Astrophysics \textbar{} Harvard \& Smithsonian, 60 Garden St., Cambridge, MA 02138, USA.}

\author[0000-0003-3654-1602]{Andrew W. Mann} 
\affiliation{Department of Physics and Astronomy, The University of North Carolina at Chapel Hill, Chapel Hill, NC 27599-3255, USA}

\author[0000-0002-0514-5538]{Luke G.~Bouma} 
\affiliation{Department of Astrophysical Sciences, Princeton University, NJ 08544, USA.}

\author[0000-0001-9911-7388]{David W.~Latham}
\affiliation{Center for Astrophysics \textbar{} Harvard \& Smithsonian, 60 Garden St., Cambridge, MA 02138, USA.}

\author{Johanna K.~Teske} 
\affiliation{Carnegie Earth and Planets Laboratory, 5241 Broad Branch Road NW, Washington, DC 20015, USA }

\author{Sharon X.~Wang} 
\affiliation{Department of Astronomy, Tsinghua University, Beijing 100084, China}

\author{Stephen A.~Shectman} 
\affiliation{The Observatories of the Carnegie Institution for Science, 813 Santa Barbara St., Pasadena, CA 91101, USA}

\author[0000-0003-1305-3761]{R.P.~Butler} 
\affiliation{Carnegie Earth and Planets Laboratory, 5241 Broad Branch Road NW, Washington, DC 20015, USA }

\author[0000-0002-5226-787X]{Jeffrey D.~Crane} 
\affiliation{The Observatories of the Carnegie Institution for Science, 813 Santa Barbara St., Pasadena, CA 91101, USA}





\begin{abstract}
Mapping the orbital obliquity distribution of young planets is one avenue towards understanding mechanisms that sculpt the architectures of planetary systems. TOI-942 is a young field star, with an age of $\mysim 60$ Myr, hosting a planetary system consisting of two transiting Neptune-sized planets in 4.3- and 10.1-day period orbits. We observed the spectroscopic transits of the inner Neptune TOI-942b to determine its projected orbital obliquity angle. Through two partial transits, we find the planet to be in a prograde orbit, with a projected obliquity angle of $|\lambda| = \postlambda$ deg. In addition, incorporating the light curve and the stellar rotation period, we find the true three-dimensional obliquity to be $\postthreed$ deg. We explored various sources of uncertainties specific to the spectroscopic transits of planets around young active stars, and showed that our reported obliquity uncertainty fully encompassed these effects. TOI-942b is one of the youngest planets to have its obliquity characterized, and one of even fewer residing in a multi-planet system. The prograde orbital geometry of TOI-942b is in line with systems of similar ages, none of which have yet been identified to be in strongly misaligned orbits. 
\end{abstract}



\section{Introduction}
\label{sec:introduction}

The characterization of a young planetary system gives us a momentary window into the process of early planet evolution. One indicator for the dynamical evolution history of planetary systems is the angle between the stellar spin axis and the orbital axes of the planets. This orbital obliquity angle is thought to be a tracer for past planetary migration events. The \emph{Kepler} mission revealed the prevalence of Neptune-sized planets around Sun-like stars \citep[e.g.][]{2011ApJ...732L..24L}, and measuring the orbital obliquities of such planets early in their history is important to mitigate confounding effects such as tidal interactions that occur over longer timescales. Close-in planets forming in their currently observed locations or migrating through the accretion disk would maintain a low obliquity, while more chaotic migration, such as planet-planet scattering, would result in more dramatic obliquity distributions \citep[see review by][]{2018ARA&A..56..175D}. 

One method to measure the obliquities of transiting planets is the Rossiter-McLaughlin effect \citep{1924ApJ....60...15R,1924ApJ....60...22M}, in which a transiting planet reduces the flux from the redshifted and blueshifted portions of a rotating star, thereby causing an apparent increase and decrease, respectively, in the observed stellar radial velocity. Though the orbital obliquities of $\sim 150$ planets have been measured to date, the sample largely consists of close-in Jovian sized planets \citep[see review by][]{2018haex.bookE...2T}.

Recent observations from the Transiting Exoplanet Survey Satellite \emph{(TESS)} revealed a system of Neptunes orbiting the star TOI-942 \citep{2021AJ....161....2Z,2020arXiv201113795C}. These studies found that although the host star cannot be placed into known moving groups or associations, it is extremely young in age, estimated to be 30-80 Myr to 1$\sigma$ uncertainties \citep{2020arXiv201113795C}, and 20-160 Myr to $3\sigma$ uncertainties \citep{2021AJ....161....2Z}. The system hosts two transiting planets, with TOI-942b residing in a 4.3-day orbit with a radius of $\mysim 4.5\,\rearth$, and TOI-942c residing in a 10.1-day orbit with a radius of $\mysim5\,\rearth$. Given the scarcity of young bright planet hosts, even at $V=11.98$ TOI-942 presented itself as a new opportunity to characterize the properties of a young planetary system. Only a dozen young transiting planets have been confirmed to date, 
and far fewer have received obliquity measurements \citep{2020ApJ...892L..21Z,2020AJ....159..112M,2020A&A...643A..25P,2020arXiv200613675A,2020A&A...641L...1M,2020ApJ...899L..13H,2020AJ....160..179M,2020AJ....160..192S}. Thus, a new measurement of a multi-planet system is a valuable addition to the effort to constrain planetary formation mechanisms.

In this letter, we present a measurement of the obliquity for TOI-942b. We obtained two partial transit spectroscopic observations using the Planet Finder Spectrograph on the 6.5\,m Magellan II telescope, used the Rossiter-McLaughlin effect to measure the projected obliquity, and modeled the system to derive a full 3D orbital obliquity for the planet. In the remaining sections, we proceed as follows: In Section \ref{sec:obs}, we describe the observational techniques used to acquire the data. In Section \ref{sec:analysis}, we describe the data analysis, as well as the consideration of possible confounding effects. In particular, we present an inferred Doppler image of the stellar surface from one of the transit observations from the iodine-free regions of the observed spectra, and though we do not directly detect the planetary shadow, we make use of it to estimate the influence of stellar activity on the Rossiter-McLaughlin effect. In Section \ref{sec:conclusion}, we present and discuss our results.

\section{Observations}
\label{sec:obs}

\subsection{6.5\,m Magellan II -- Planet Finder Spectrograph}
\label{sec:pfs}

Transits of TOI-942b were observed with the Planet Finder Spectrograph \citep[PFS,][]{2006SPIE.6269E..31C,2008SPIE.7014E..79C,2010SPIE.7735E..53C} on the 6.5\,m Magellan II (Clay) Telescope, located at Las Campanas Observatory, Chile. PFS is a high resolution slit fed echelle spectrograph covering the wavelength range of $3910$--$7340\,\AA$. Our observations utilized the $0.3\arcsec$ slit, with the detector sampled in $3\times3$ binning mode, with the corresponding effective spectral resolving power of $R\approx 110,000$. The transit sequence was observed through the iodine gas cell to allow for modeling of the instrument profile, achieving $\sim 5\ms$ precision velocities in this mode. 

We obtained two partial transits of TOI-942b on the nights of 2020-10-28 and 2021-01-01 UTC. A total of seven observations were obtained on 2020-10-28, covering 1.0 hours pre-ingress to 1.0 hours post-ingress, with the sequence cut short due to poor weather. Twelve observations were obtained on 2021-01-01, spanning most of the transit, from 0.3 hours post-ingress to 0.7 hours post-egress. Each observation was obtained with an integration time of 1200s. An iodine-free template observation was obtained on the night of 2020-12-27 UTC. 

The velocities were derived from each observation using a custom IDL pipeline as per \citet{1992PASP..104..270M} and \citet{1996ApJ...464L.153B}. The velocities are shown in Figure~\ref{fig:velmodel}. In addition, we made use of the iodine-free template to derive a rotational broadening velocity of $\vsini=14.26\pm0.5\kms$ and macroturbulent broadening $v_\mathrm{mac}=4.1 \pm 0.5\,\kms$ for TOI-942, measured from a fit to the line profiles via a least-squares deconvolution of the observation against a non-rotating spectral template \citep{1997MNRAS.291..658D}. 

\subsection{Updated \emph{TESS} observations}
\label{sec:TESS}

TOI-942 received continuous monitoring from \emph{TESS} over two separate sectors of observations. Discovery light curves were obtained during the primary \emph{TESS} mission in Sector 5, between 2018 November 15 and December 11. These observations were obtained in the 30-minute cadence full frame images, and light curves were extracted via the MIT quicklook pipeline \citep{2020RNAAS...4..206H}. 

\begin{figure*}
    \centering
    \includegraphics[width=0.8\linewidth]{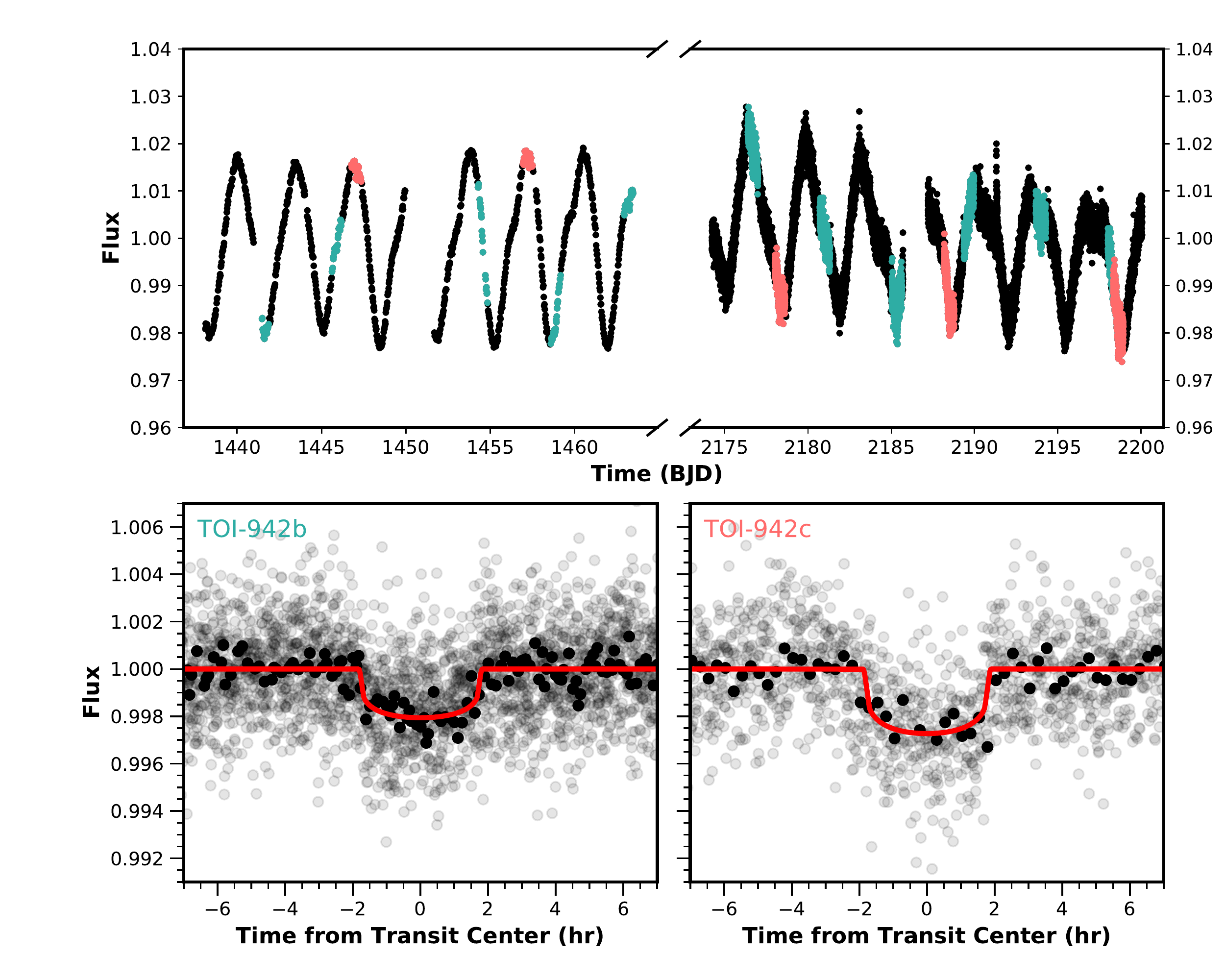}
    \caption{\emph{TESS} photometry of the TOI-942 system. \textbf{Top:} \emph{TESS} light curves of TOI-942, with Sector 5 full frame images at 30-minute cadence on the left and Sector 32 2-minute cadence observations on the right. Data points corresponding to a transit of TOI-942b are shaded in light blue, while those corresponding to a transit of TOI-942c are in pink. \textbf{Bottom left:} Detrended, phase folded transits of TOI-942b. 30-minute data points are in black, while 2-minute data points are in grey. The best-fit short cadence model is in red. \textbf{Bottom right:} Detrended and phase folded transits of TOI-942c, along with the best-fit model.}
    \label{fig:lightcurve}
\end{figure*}

TOI-942 was also observed during Sector 32 of the extended \emph{TESS} mission, between 2020 September 22 and October 21. Sector 32 observations were obtained at 2-minute cadence via target pixel stamps, with light curves made available through the Science Processing Observation Center \citep[SPOC,][]{2016SPIE.9913E..3EJ} analysis and via the Mikulski Archive for Space Telescopes (MAST). As these observations were obtained after the publication of the discovery papers, they are shown in Figure~\ref{fig:lightcurve}.

In particular, we note the presence of at least two flare events during the extended mission observations. Flares were not seen during the primary mission longer cadence observations, and would have been detectable in the long cadence data should they have occurred. Frequent flares are expected for late-type stars exhibiting significant stellar activity, which further corroborates the youth of the system. 

\section{Analysis and Modeling}
\label{sec:analysis}

We performed a joint modeling of the Rossiter-McLaughlin observations and all available \emph{TESS} photometric observations for both TOI-942b and TOI-942c. The in-transit velocities were modeled as per the \citet{2011ApJ...742...69H} approximation. The photometric transits were computed using the \citet{2002ApJ...580L.171M} model, implemented via BATMAN \citep{2015PASP..127.1161K}. The free parameters for the light curve model for each planet are the times of transit center $t_0$, the orbital periods $P$, the planetary radius ratios $R_p/R_\star$, and the eccentricity parameters $\sqrt{e}\cos{\omega}$ and $\sqrt{e}\sin{\omega}$, where $e$ is the eccentricity and $\omega$ is the longitude of periastron. Specifically for the Rossiter-McLaughlin model of TOI-942b, there is also the projected orbital obliquity angle $\lambda$, as well as the host star rotational broadening $\vsini$ and the macroturbulent broadening $v_\mathrm{mac}$. We employed the use of a jitter term to characterize and account for overall uncertainty due to stochastic stellar noise. A linear polynomial was fit to the first partial transit on 2020-10-28, and a quadratic polynomial to the second on 2021-01-01, to correct for the hours-timescale effects of stellar activity on the observed velocities. Because of the low number of data points for the first transit, we found too high a degree of degeneracy if a quadratic polynomial were to be used.

The stellar mass and radius were simultaneously modeled via the MIST isochrones \citep{2016ApJS..222....8D}, and constrained by photometric magnitudes and parallax priors from \emph{Gaia} \emph{G}, {Bp}, {Rp} \citep{2018A&A...616A...1G}, HIPPARCHUS THYCHO $B$ and $V$ bands \citep{1997AA...323L..49P}, 2MASS $J$, $H$, and $Ks$ bands \citep{2006AJ....131.1163S}. We also apply a Gaussian prior of $50\pm30$ Myr for the age of TOI-942 to the fitting to help restrict the parameter space of the isochrone interpolation. In addition, Gaussian priors, derived from the spectrum in Section~\ref{sec:obs}, were adopted for $\vsini$ and $v_\mathrm{mac}$, and the limb darkening coefficients were held fixed to their theoretically interpolated values \citep{Claret:2011,2017A&A...600A..30C}. The \emph{TESS} photometric transits and the stellar isochrones co-constrain the critical transit parameters to help better propagate the uncertainties of the system, especially for the $t_0,$ $P,$ $R_p,$ $R_\star$, $M_\star,$ and $i$ parameters. This ensures that we encompass the critical sources of uncertainty in our obliquity measurement for TOI-942b. 

\begin{figure}
    \centering
    \includegraphics[width=\linewidth]{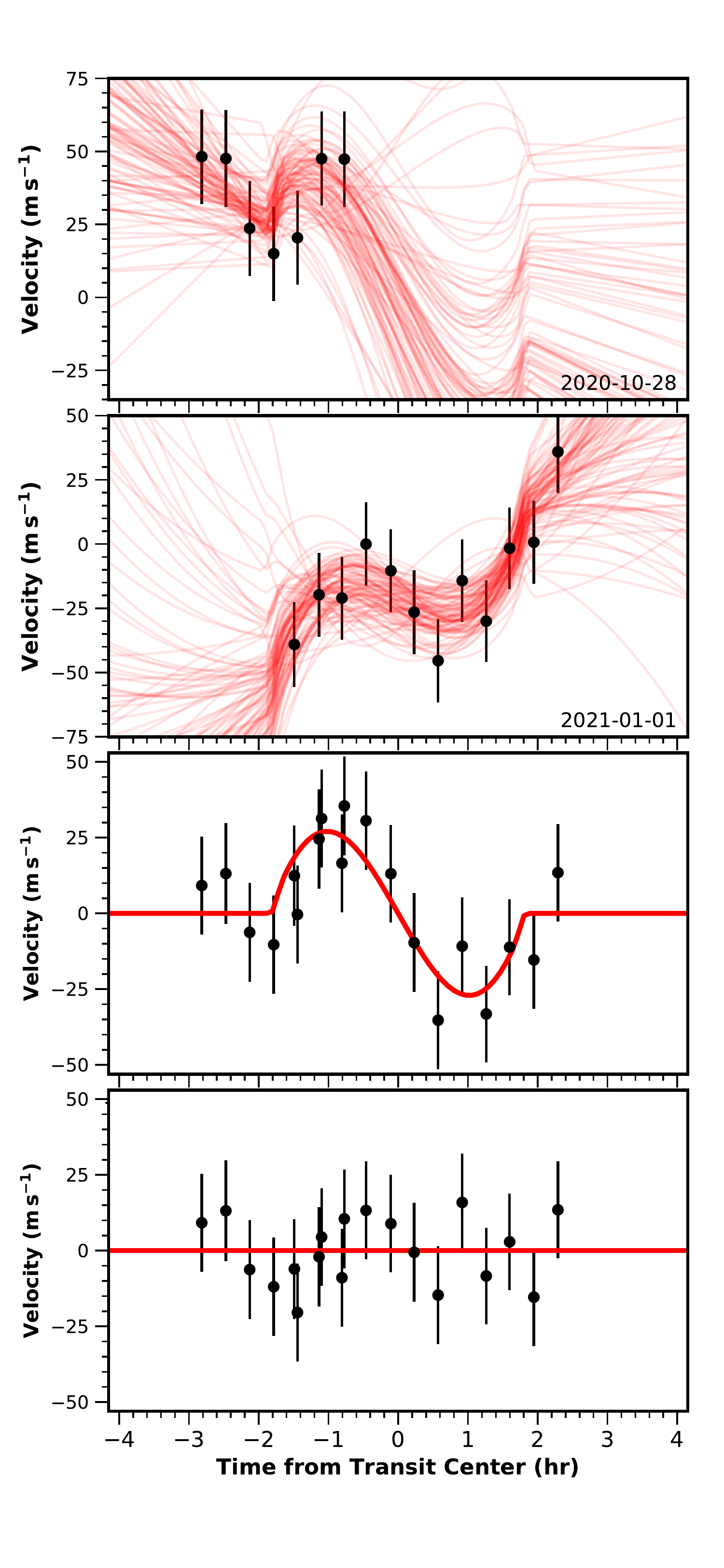}
    \caption{Radial velocities during the transits of TOI-942b as measured by PFS, with error bars incorporating the measurement uncertainties and the fitted radial velocity jitter.  \textbf{The data behind this figure is available on the online version of this publication.} \textbf{Top}: Observations from 2020-10-28. Samples are drawn from the posterior and the corresponding models shown in red to demonstrate the constraints that the dataset places on the resulting Rossiter-McLaughlin models. \textbf{Top Middle}: Observations from 2021-01-01, with models drawn from the posterior shown in red. \textbf{Bottom Middle}: Detrended combined velocity data for both observations (black) and the best fit model for the Rossiter McLaughlin effect (red). \textbf{Bottom}: Residual plot for the best-fit model in the panel above}
    \label{fig:velmodel}
\end{figure}

We perform the global modeling using a Monte Carlo Markov Chain (MCMC) analysis, implemented via the ensemble sampler \emph{emcee}  \citep{2013PASP..125..306F}. The results are shown in Table \ref{tab:planetparam}. Typical draws from the posterior and the best-fit velocity curve are shown in Figure \ref{fig:velmodel}, and the best-fit light curve is shown in Figure \ref{fig:lightcurve}.

Finally, we make use of the prominent rotational signal seen in the light curves to derive the 3D obliquity angle of TOI-942b. Following \citet{2020AJ....159...81M}, and adopting a rotation period of $3.40\pm0.37$\,days from \citet{2021AJ....161....2Z}, we find the stellar inclination $I_\star$ to be constrained to $> 72.5^\circ$ at $1\sigma$ significance. Together with the projected obliquity measured from the Rossiter-McLaughlin effect, we find an overall 3D obliquity for the orbital plane of TOI-942b to be $\postthreed$ degrees.


\begin{deluxetable*}{lrrrr}
\tablewidth{0pc}
\tabletypesize{\scriptsize}
\tablecaption{
    Derived parameters for TOI-942b
    \label{tab:planetparam}
}
\tablehead{ \\
    \multicolumn{1}{c}{~~~~~~~~Parameter~~~~~~~~}   &
    \multicolumn{1}{c}{Joint model} &
    \multicolumn{1}{c}{Priors} &
}
\startdata
\sidehead{\textbf{TOI-942b}}
~~~$T_0$ (BJD)              \dotfill & $2458441.5790_{-0.0021}^{+0.0021}$ & Uniform\\
~~~$P$ (days)             \dotfill    &  $ 4.32421_{-0.000019 }^{+0.000019 }$ & Uniform \\
~~~$\rpl/\rstar$ \dotfill & $0.03995_{-0.00063}^{+0.00065}$ & Uniform\\
~~~$i$ (deg) \dotfill & $89.966_{-0.024}^{+0.020}$ & Uniform\\
~~~$|\lambda|$ (deg)      \dotfill    &  $\postlambda$ & Uniform\\
~~~$\sqrt{e}\cos{\omega}$      \dotfill    &  $-0.564_{-0.031}^{+0.032}$ & Uniform\\
~~~$\sqrt{e}\sin{\omega}$      \dotfill    &  $0.16_{-0.43}^{+0.30}$ & Uniform\\
~~~$e$      \dotfill    &  $0.34_{-0.14}^{+0.10}$ & Derived\\
~~~$\omega$ (deg)    \dotfill    &  $-16_{-40}^{+28}$ & Derived\\
~~~3D obliquity (deg) \dotfill & $\postthreed$ & Derived\\
~~~$\rpl$ ($R_\oplus$)       \dotfill    & $3.89_{-0.23}^{+0.25}$ & Derived\\
~~~$a$ ($AU$) \dotfill & $0.04866_{-0.00013}^{+0.00016}$ & Derived\\
\sidehead{\textbf{TOI-942c}}
~~~$T_0$ (BJD)              \dotfill & $2458447.0563_{-0.0020}^{+0.0019}$ & Uniform\\
~~~$P$ (days)             \dotfill    &  $10.156272_{-0.000034}^{+0.000037}$ & Uniform \\
~~~$\rpl/\rstar$ \dotfill & $0.0479_{-0.0012}^{+0.0018}$ & Uniform\\
~~~$i$ (deg) \dotfill & $89.16_{-0.30}^{+0.38}$ & Uniform\\
~~~$\sqrt{e}\cos{\omega}$      \dotfill    &  $-0.535_{-0.109}^{+0.077}$ & Uniform\\
~~~$\sqrt{e}\sin{\omega}$      \dotfill    &  $-0.18_{-0.30}^{+0.59}$ & Uniform\\
~~~$e$      \dotfill    &  $0.32_{-0.16}^{+0.23}$ & Derived\\
~~~$\omega$ (deg)    \dotfill    &  $20_{-33}^{+63}$ & Derived\\
~~~$\rpl$ ($R_\oplus$)       \dotfill    & $4.67_{-0.30}^{+0.34}$ & Derived\\
~~~$a$ ($AU$) \dotfill & $0.08598_{-0.00022}^{+0.00027}$ & Derived\\
\sidehead{\textbf{Stellar}}
~~~Mass ($M_{\odot}$) \dotfill & $0.8220_{-0.0064}^{+0.0079}$ & Uniform\\
~~~Radius ($R_{\odot}$) \dotfill & $0.894_{-0.052}^{+0.056}$ & Uniform\\
~~~Rotational broadening $\vsini$ (m/s) \dotfill & $14240_{-530}^{+510}$ & $\mathcal{G}(14260,500)$\\
~~~Macroturbulence $v_\mathrm{mac}$ (m/s) \dotfill & $4050_{-470}^{+490}$ & $\mathcal{G}(4100,500)$\\
~~~Radial velocity jitter  (m/s) \dotfill & $15.1_{-4.2}^{+6.2}$ & Uniform \\
~~~Age (Myr)\dotfill & $53_{-21}^{+22}$ & $\mathcal{G}(50,30)$\\
~~~Parallax (mas) \dotfill & $6.605_{-0.014}^{+0.012}$ & $\mathcal{G}(6.6,0.015)$\\
~~~Limb darkening coefficients (R-M) \dotfill & (0.5751, 0.1961)& Fixed\\
~~~Limb darkening coefficients (TESS) \dotfill & (0.4006, 0.2243)& Fixed\\
\sidehead{\textbf{Polynomial Coefficients\tablenotemark{a}}}
~~~Transit 1 Linear ($b_1$) \dotfill & $-1400_{-1000}^{+1000}$ & Uniform\\
~~~Transit 1 Constant ($c_1$) \dotfill & $2_{-20}^{+20}$ & Uniform\\
~~~Transit 2 Quadratic ($a_2$)\dotfill & $-3000_{-40000}^{+48000}$ & Uniform\\
~~~Transit 2 Linear ($b_2$)\dotfill & $2060_{-790}^{+600}$ & Uniform\\
~~~Transit 2 Constant ($c_2)$\dotfill & $-21.4_{-7.4}^{+7.0}$ & Uniform\\
\sidehead{\textbf{Spot Modeling Parameters}}
~~~Spot 1 Radius (deg)  \dotfill & $24_{-10}^{+9}$ & Uniform\\
~~~Spot 1 Contrast\tablenotemark{b}  \dotfill & $1.22_{-0.13}^{+0.24}$ & Uniform\\
~~~Spot 1 Latitude (deg) \dotfill & $46_{-35}^{+7}$ & Uniform\\
~~~Spot 1 Longitude (deg) \tablenotemark{c} \dotfill & $-60_{-30}^{+13}$ & Uniform\\
~~~Spot 2 Radius  (deg) \dotfill & $6_{-2}^{+5}$ & Uniform\\
~~~Spot 2 Contrast  \dotfill & $1.145_{-0.050}^{+0.092}$ & Uniform\\
~~~Spot 2 Latitude (deg) \dotfill & $3_{-7}^{+13}$ & Uniform\\
~~~Spot 2 Longitude (deg) \dotfill & $-2.7_{-1.2}^{+1.9}$ & Uniform\\
~~~Spot 3 Radius (deg)  \dotfill & $15_{-8}^{+9}$ & Uniform\\
~~~Spot 3 Contrast  \dotfill & $1.19_{-0.10}^{+0.23}$ & Uniform\\
~~~Spot 3 Latitude (deg) \dotfill & $49_{-42}^{+5}$ & Uniform\\
~~~Spot 3 Longitude (deg) \dotfill & $72_{-21}^{+14}$ & Uniform\\
\enddata
\tablenotetext{a}{(2020-10-28 Transit RV)$=b_1$(Phase)$+c_1$\\(2021-01-01 Transit RV)$=a_2$(Phase)$^2+b_2$(Phase)$+c_2$}
\tablenotetext{b}{Inferred contrast indicate bright surface features.}
\tablenotetext{c}{The meridian at the start of the observations is defined as having $0^\circ$ longitude.}
\end{deluxetable*}

We note that our stellar and transit modeling using the joint Sector-5 and 32 observations suggest non-circular orbits for TOI-942b and c. The possibility of non-circular orbits was originally raised by the discovery papers \citep{2020arXiv201113795C,2021AJ....161....2Z}, inferred via analyses of the original single-sector 30-minute cadenced observations. We compared the transit density resulting from our global fit incorporating stellar models to that resulting from a fit forcing circular orbits, and found them to be inconsistent at the $3\sigma$ level. We note that inferences from the photo-eccentric effect depend strongly on our understanding of the stellar properties. We derive a stellar mass of $0.8220_{-0.0064}^{+0.0079} M_\odot$ and radius of $0.894_{-0.052}^{+0.056} R_\odot$ for TOI-942, based on its spectral energy distribution, age constraints described in the discovery papers, and the MIST isochrones. To check these results, we followed additional techniques to check our stellar parameters independent of the transit light curve. We followed the $M_K$ relationship in \citet{2019ApJ...871...63M} to derive a stellar radius of $0.89\pm0.04$, fully consistent with our analysis. Similarly, we also modeled the spectral energy distribution with interpolated isochrones using the EXOFASTv2 suite \citep{2019arXiv190709480E}, finding TOI-942 to have a mass of $0.831_{-0.025}^{+0.035}\,M_\odot$ and radius of $0.876_{-0.024}^{+0.026}\,R_\odot$, fully consistent with our global model fit.  Future works can explore the possible scenarios that might have led to the eccentric orbits of these young Neptunes.


\subsection{The influence of correlated noise}
\label{sec:systematic_noise}

Short term correlated noise in the radial velocities has the potential to greatly affect our obliquity measurement. These can arise from spot crossing events that were temporally unresolved, granulation, or possible terrestrial atmospheric and instrumental variations unaccounted for in our uncertainty estimates. We performed a number of tests to examine for the effects of such noise within our data set on our results. We took the residuals from our best-fit model of the Rossiter-McLaughlin effect, cyclically permuted them among the data points of each transit, and added them back to the best-fit model to create a new velocity dataset. With each iteration, we performed the same MCMC analysis as we did on the original data. We repeated this for all possible permutations of the residuals, each of which gave us a different best-fit value of $\lambda$. We found a mean obliquity value of $16^\circ$, with a scatter of $32^\circ$ about the mean, from this experiment. The scatter is less than the reported posterior uncertainty in $\lambda$ of $41^\circ$. As such, we conclude that our reported obliquity accounts for the presence of short term stellar activity via the jitter term. Beyond short term correlated noise, we also explore any systematic biases that may exist in our derived obliquity measurements in the discussion below.

\subsection{The influence of spots and active regions in systematically biasing the Rossiter-McLaughlin effect}
\label{sec:spotmodel}

The presence of stellar activity and star spots can induce systematic biases in the Rossiter-McLaughlin velocity anomaly we observe. Standard models of the Rossiter-McLaughlin effect assume a radially symmetric surface brightness distribution for the host star. Young stars, however, exhibit significant stellar activity in their light curves and spectroscopic time series, and are thought to have stellar surfaces with large star spots and active regions. The \emph{TESS} light curves of TOI-942 exhibits spot modulated rotational variability at the $\mysim4$\% level, and subsequent radial velocity follow-up found the star to exhibit long term jitter variability at the 65~\ms level \citep{2020arXiv201113795C}. When planets occult these bright and dark regions, they induce bumps in the photometric transits and Rossiter-McLaughlin velocities we measure. These bumps in velocity can have an effect on the derived projected obliquities \citep[e.g.][]{2013A&A...549A..35O,2018A&A...619A.150O}. 

\begin{figure*}
    \centering
        \begin{tabular}{cc}
                 \includegraphics[width=0.5\linewidth]{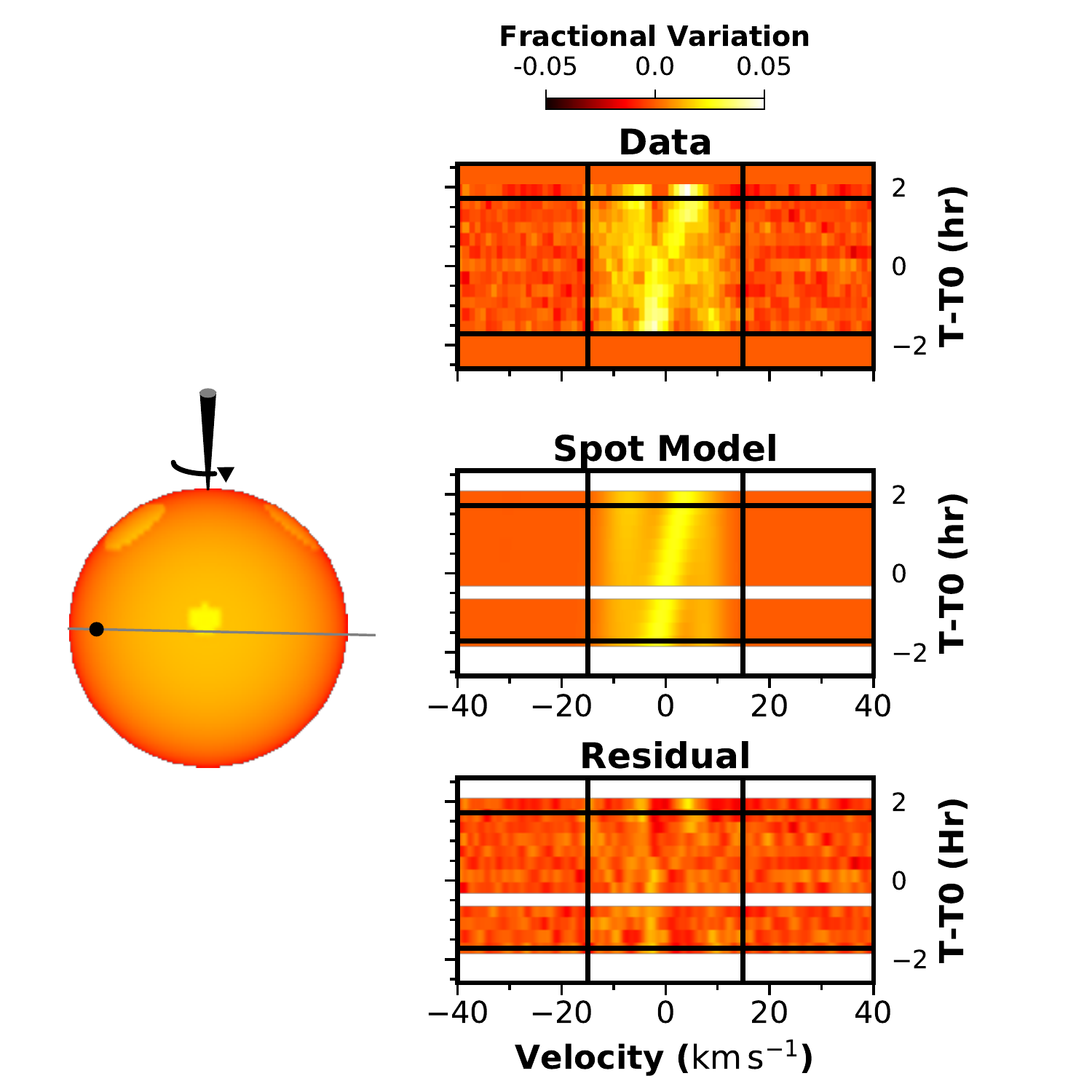} &
                 \includegraphics[width=0.4\linewidth]{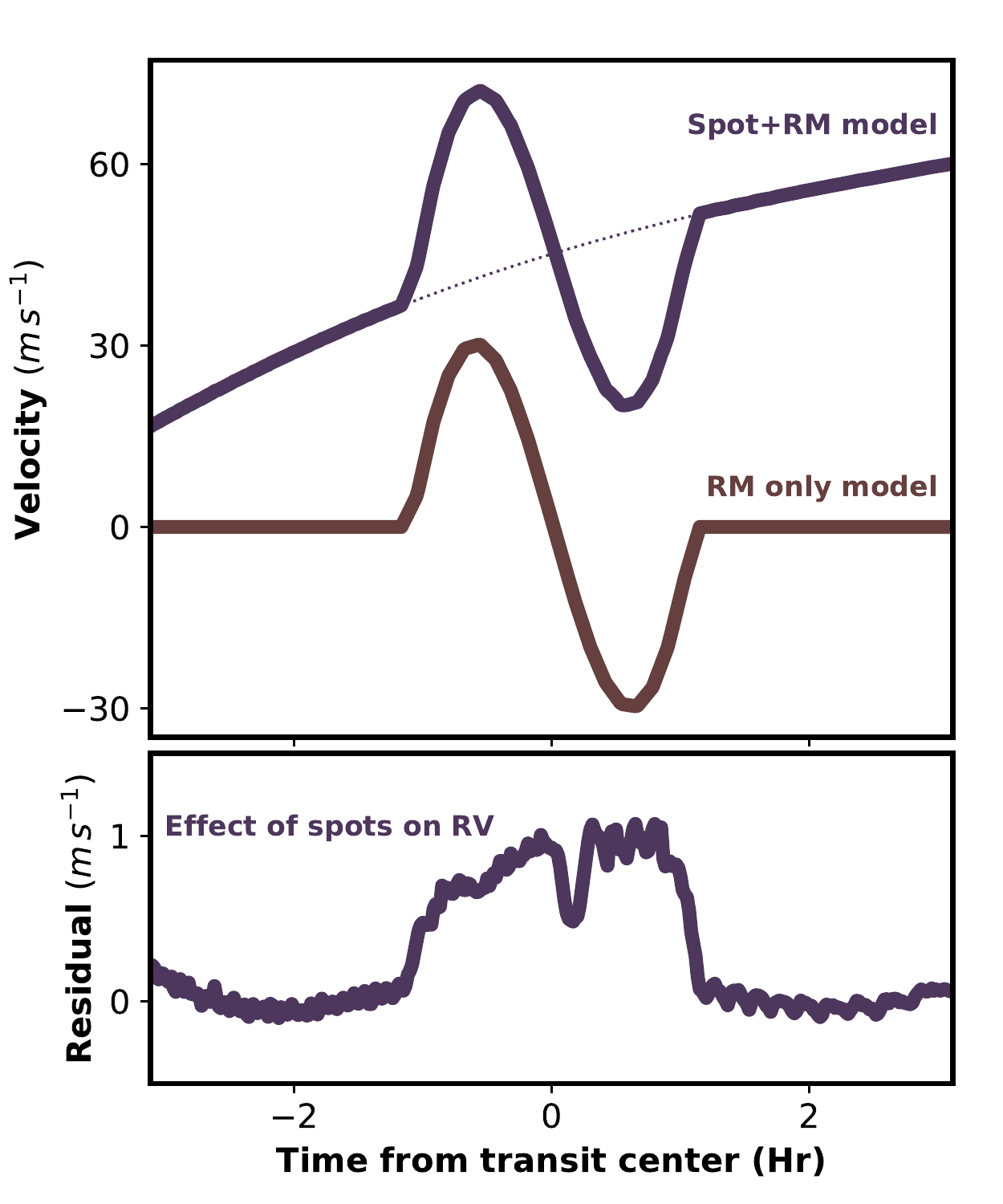} \\
        \end{tabular}
    \caption{Our model interpretation for the spot configuration and resulting Rossiter-McLaughlin model for the transit of TOI-942b on 2021-01-01. Note that the shadow cast by the planet is not visible at the signal-to-noise of this observation, and is not detectable in this figure. \textbf{Left:} Illustration of the modeled stellar surface. The path of the transiting planet is marked by the grey line. \textbf{Top center:} The line profile residuals from which this model is derived, plotted as a color map as a function of velocity and orbital phase. The diagonal stripes are interpreted as bright spots traversing the stellar surface. \textbf{Center:} Model of the line profile residuals from the spot model. \textbf{Bottom center:} \textbf{The line profile residuals after the subtraction of the best fit spot model.} \textbf{Top right:} The model Rossiter-McLaughlin curve expected for this star spot configuration. A long-term trend is induced by the rotating spotted star. We also model a hypothetical spot crossing event occurring $\mysim30$ minutes after mid-transit \textbf{Bottom right:} The isolated effects of the star spots on the Rossiter-McLaughlin curve. The long term trend is removed with a polynomial. The star spot crossing event induces a 0.5\ms\ level bump on the observed velocities. The uneven illumination of the star induces a bias on the in-transit velocities at the $<0.5\,\ms$ level. Note that the PFS velocities we obtained have per-point uncertainties of $\sim8\ms$, so neither effect would have been detectable, or strong enough to influence our analysis.}
    \label{fig:spot_dt}
\end{figure*}

To estimate the magnitude of the influence of stellar activity on the observed Rossiter-McLaughlin velocities of TOI-942, we can create models of the stellar surface from the PFS spectral line profiles. We derive line broadening profiles via a Least-Squares Deconvolution (LSD) \citep{1997MNRAS.291..658D} of the iodine-free regions of the PFS spectra (4000-4900\,\AA). The deconvolution is performed against a non-rotating synthetic spectrum of TOI-942, generated using the ATLAS9 model atmospheres with stellar parameters matching that of TOI-942 \citep{Castelli:2004}. We averaged all the line profiles measured using PFS on the night of 2021-01-01 to form a master line profile. From which, we subtract each individual observation to visualize the line profile variations over the course of the $\sim 4$ hours observations on 2021-01-01. The line profile variations are shown as a function of time in Figure~\ref{fig:spot_dt}. 

We model the line profile variations by creating a model of the spot distribution on the stellar surface. The full model and its free parameters are described in \citet{2020ApJ...892L..21Z}. Briefly, we generate a set of circular spots, with free parameters describing each spot's radius, contrast, longitude, and latitude, about the stellar surface. We then model the influence of these spots on the line profiles, and match them against the observations. The model is compared via a series of MCMC analyses (using \emph{emcee};  \citealt{2013PASP..125..306F}), with the number of spots required determined by an examination of the Bayesian Information Criteria of the fit. 

We found the 2021-01-01 observations can be fit with a set of three bright regions on the stellar surface of TOI-942, each with radii $\mysim15^\circ$, and contrast of $\mysim1.2$ with respect to the unspotted stellar surface. This spot configuration, and the modeled line profile variations, are shown in Figure~\ref{fig:spot_dt}, and the spot parameters shown in Table~\ref{tab:planetparam}. This configuration includes two spots near the polar regions of the star, responsible for the near-vertical bright trails seen in the Doppler tomographic mapping. The model also includes a single near-equatorial spot responsible for the diagonal bright trail. 

Note that the inferred influence of the star spots on the line profiles is more than an order of magnitude larger than that expected to be induced by the transit of TOI-942b. At the noise-level of these observations, the actual planetary transit was not recovered in the line profile residuals. The planet’s shadow has an expected depth of 0.6\%, the spot signal has a depth of $\sim$3\%, while the per-exposure uncertainties are at the 0.9\% level. As such, even if the star lacked any activity, we would still have had trouble identifying the transiting planet from the iodine-free regions of the PFS observations. 

Figure~\ref{fig:spot_dt} shows the modeled Rossiter-McLaughlin effect for this star spot model. The top panel shows the Rossiter-McLaughlin model, with a smoothly varying trend at the 10\,\ms\ level due to the rotating spotted star. This trend can be satisfactorily removed at the $\ll 1\,\ms$ level by simultaneously fitted a low order polynomial to the velocities.  From our observations, a trend of similar magnitude was observed on both nights, and its effect accounted for via a linear and quadratic model (Section~\ref{sec:analysis}). 

The unspotted Rossiter-McLaughlin model is plotted for comparison. When we subtract a polynomial from this spotted Rossiter-McLaughlin model, and then remove the unspotted model, the resulting residuals isolate the effect of spots on the transit velocities. The bottom right panel of Figure~\ref{fig:spot_dt} shows the magnitude of these effects. 

Within the spot model, there is a degeneracy for spots being present in the Northern or Southern hemispheres of the star. We chose to place our spots in the same hemisphere as the planetary transit, so to contrive a scenario where the planet crosses directly infront of one of the spots. This model configuration allows us to test the effects of such 'spot-crossing' events would have on our Rossiter-McLaughlin observations. We do not know if spot crossing events occurred during either of our observations. 

Our simulations reveal that a spot crossing induces a $\sim0.5\,\ms$ level sharp bump. The simulations also show another subtle $\mysim0.5\,\ms$ effect throughout the transit, which is due to one hemisphere of the star being brighter than the other as a result of the spot configuration. Our per-point uncertainty is at the 8\,\ms~level, and as such the effect of the spot crossing and the uneven illumination are too small to influence our eventual analysis of the Rossiter-McLaughlin effect. 

We note that many degeneracies exist when interpreting Doppler imaging results, and the model presented is only one interpretation of the observed line profile variations, presented to test the effects of star spots on the interpretation of the Rossiter-McLaughlin observation. In particular, we choose the minimal number of spots that may satisfactorily fit the line profile variations, but the true configuration likely involves numerous additional features. However, as the number of spots increase, their respective sizes and contrasts should decrease, and their distribution across the stellar surface should become more uniform, as to sustain the same levels of photometric and spectroscopic variations. As such, transits across a larger number of smaller spots should yield similar, or smaller deviations. The interpretation we present here should be taken as an upper limit on influence of stellar spots on the Rossiter-McLaughlin model. In addition, we also note that Doppler imaging has been shown to infer bright spots, as is the case here, while interferometric images have shown otherwise. As such, the true spot configuration may be very different to those presented here \citep[e.g][]{2017ApJ...849..120R}.


\section{Conclusion}
\label{sec:conclusion}

TOI-942b is among the youngest planets to have its orbital obliquity measured, and one of only a few planets in multi-planet systems with obliquity measurements. Using two PFS partial transits over the course of a few months, we observed the Rossiter-McLaughlin effect and measured a projected obliquity $(|\lambda|)$ for TOI-942b to be within 50 degrees of its host star's plane of rotation. Further incorporation of the host rotation period allows us to place a limit on the true 3D obliquity of the orbital plane of TOI-942b, finding it to be prograde to within $50^\circ$.

Most significantly, we can likely reject the possibility of dramatic dynamical interactions between the orbiting planets playing a role in the past history of the TOI-942 system, as these would have likely resulted in misalignment. We can also rule out the mechanisms that resulted in oblique orbits for compact planetary systems, such as Kepler-56 \citep{2013Sci...342..331H} and K2-290 \citep{2021arXiv210207677H}. Kepler-56 is a red giant hosting two coplanar misaligned planets in 10.5- and 21.4-day orbits, and radial velocity measurements of the system revealed a third companion that likely provided a torque on the planets shortly after their formation. The K2-290 system contains two coplanar planets, these ones in 9.2- and 48.4-day orbits, believed to have been formed from a protoplanetary disk tilted by a wide stellar companion. These are mechanisms acting on the Myr timescale \citep{2017MNRAS.468..549B,2017AJ....153..210H}, in the presence of external companions, and would have tilted TOI-942 within its $\mysim 60$\,Myr lifetime. 


 \begin{figure}
     \centering
     \includegraphics[width=\linewidth]{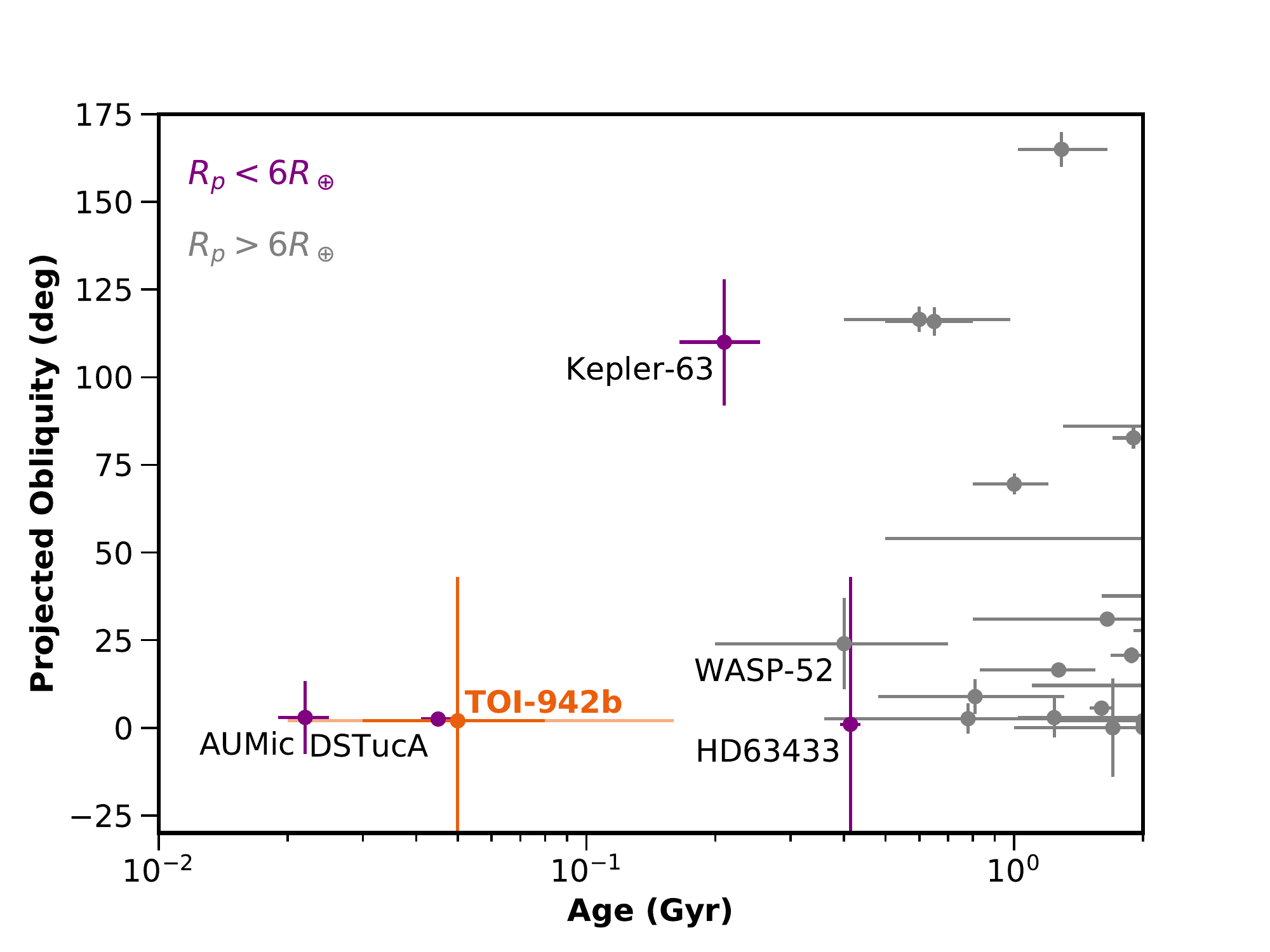}
     \caption{The distribution of planets for which there are both age estimates and obliquity measurements. Planets younger than 500 Myr are labeled, and the 3$\sigma$ age range for TOI-942 is shown in a lighter orange. Planets smaller than $6R_{\earth}$ are shown in purple, while larger planets are shown in gray. Note the rarity of misaligned planets around young stars measured to date. }
     \label{fig:age_obliq}
 \end{figure}
 
Figure \ref{fig:age_obliq} shows all of the planets for which we have both age estimates and obliquity measurements. We can see that there are no strongly misaligned systems younger than 100 Myr. These include AU Mic b \citep{2020A&A...643A..25P,2020A&A...641L...1M,2020ApJ...899L..13H,2020arXiv200613675A}, DS Tuc Ab \citep{2020ApJ...892L..21Z,2020AJ....159..112M}, and HD63433 b \citep{2020AJ....160..179M}. We also note that recent observations of 24 Myr old $\beta$ Pictoris b measured with VLTI/GRAVITY spectro-interferometry showed that the planet is well aligned to the disk \citep{2020ApJ...897L...8K}, though not marked in Figure~\ref{fig:age_obliq} due to significant differences between $\beta$ Pictoris b and the other systems investigated. 

While the youngest small planets are so far found to be in prograde orbits, there is evidence for misalignment within the broader population of small planets. \citet{2021AJ....161..119R} demonstrated that misaligned small planets $(M_p<100\,\mearth)$ around cool stars $(\teff < 6250\,K)$ are commonplace, with 8 of the 19 such systems investigated so far exhibiting orbital obliquities inclined to $>30^\circ$. With mechanisms such as disk dispersal, Kozai-Lidov oscillations, and nodal precession all occurring on extremely fast timescales ($\mysim10^5 - 10^6$ years), the current lack of misalignment among the youngest small planets is noteworthy. Further investigation of these planets around young stars will help test the timescales of these tilting mechanisms, and help explain the origins of misaligned Neptune-sized planets around mature stars \citep[e.g.][]{2011ApJ...743...61S,2018Natur.553..477B,2018AJ....155..255Y,2019AJ....157..137K}.

\acknowledgements  
CW and GZ thank the support of the TESS Guest Investigator Program G03007.
This research has made use of the NASA Exoplanet
Archive, which is operated by the California Institute of Technology,
under contract with the National Aeronautics and Space Administration
under the Exoplanet Exploration Program. 
Funding for the TESS mission is provided by NASA's Science Mission directorate. We acknowledge the use of public TESS Alert data from pipelines at the TESS Science Office and at the TESS Science Processing Operations Center. This research has made use of the Exoplanet Follow-up Observation Program website, which is operated by the California Institute of Technology, under contract with the National Aeronautics and Space Administration under the Exoplanet Exploration Program. This paper includes data collected by the TESS mission, which are publicly available from the Mikulski Archive for Space Telescopes (MAST).
Resources supporting this work were provided by the NASA High-End Computing (HEC) Program through the NASA Advanced Supercomputing (NAS) Division at Ames Research Center for the production of the SPOC data products.

\facility{Magellan, TESS, Exoplanet Archive}
\software{emcee \citep{2013PASP..125..306F}, batman \citep{2015PASP..127.1161K}}
\bibliographystyle{apj}
\bibliography{refs}

\begin{thebibliography}{}
\expandafter\ifx\csname natexlab\endcsname\relax\def\natexlab#1{#1}\fi

\bibitem[{{Addison} {et~al.}(2020){Addison}, {Horner}, {Wittenmyer},
  {Plavchan}, {Wright}, {Nicholson}, {Marshall}, {Clark}, {Kane}, {Hirano},
  {Kielkopf}, {Shporer}, {Tinney}, {Zhang}, {Ballard}, {Bedding}, {Bowler},
  {Mengel}, {Okumura}, \& {Gaidos}}]{2020arXiv200613675A}
{Addison}, B.~C., {Horner}, J., {Wittenmyer}, R.~A., {et~al.} 2020, arXiv
  e-prints, arXiv:2006.13675

\bibitem[{{Becker} \& {Adams}(2017)}]{2017MNRAS.468..549B}
{Becker}, J.~C., \& {Adams}, F.~C. 2017, \mnras, 468, 549

\bibitem[{{Bourrier} {et~al.}(2018){Bourrier}, {Lovis}, {Beust}, {Ehrenreich},
  {Henry}, {Astudillo-Defru}, {Allart}, {Bonfils}, {S{\'e}gransan}, {Delfosse},
  {Cegla}, {Wyttenbach}, {Heng}, {Lavie}, \& {Pepe}}]{2018Natur.553..477B}
{Bourrier}, V., {Lovis}, C., {Beust}, H., {et~al.} 2018, \nat, 553, 477

\bibitem[{{Butler} \& {Marcy}(1996)}]{1996ApJ...464L.153B}
{Butler}, R.~P., \& {Marcy}, G.~W. 1996, \apjl, 464, L153

\bibitem[{{Carleo} {et~al.}(2021){Carleo}, {Desidera}, {Nardiello},
  {Malavolta}, {Lanza}, {Livingston}, {Locci}, {Marzari}, {Messina}, {Turrini},
  {Baratella}, {Borsa}, {D'Orazi}, {Nascimbeni}, {Pinamonti}, {Rainer}, {Alei},
  {Bignamini}, {Gratton}, {Micela}, {Montalto}, {Sozzetti}, {Squicciarini},
  {Affer}, {Benatti}, {Biazzo}, {Bonomo}, {Claudi}, {Cosentino}, {Covino},
  {Damasso}, {Esposito}, {Fiorenzano}, {Frustagli}, {Giacobbe}, {Harutyunyan},
  {Leto}, {Magazz{\`u}}, {Maggio}, {Mainella}, {Maldonado}, {Mallonn},
  {Mancini}, {Molinari}, {Molinaro}, {Pagano}, {Pedani}, {Piotto}, {Poretti},
  {Redfield}, \& {Scandariato}}]{2020arXiv201113795C}
{Carleo}, I., {Desidera}, S., {Nardiello}, D., {et~al.} 2021, \aap, 645, A71

\bibitem[{{Castelli} \& {Kurucz}(2004)}]{Castelli:2004}
{Castelli}, F., \& {Kurucz}, R.~L. 2004, ArXiv Astrophysics e-prints,
  astro-ph/0405087

\bibitem[{{Claret}(2017)}]{2017A&A...600A..30C}
{Claret}, A. 2017, \aap, 600, A30

\bibitem[{{Claret} \& {Bloemen}(2011)}]{Claret:2011}
{Claret}, A., \& {Bloemen}, S. 2011, \aap, 529, A75

\bibitem[{{Crane} {et~al.}(2006){Crane}, {Shectman}, \&
  {Butler}}]{2006SPIE.6269E..31C}
{Crane}, J.~D., {Shectman}, S.~A., \& {Butler}, R.~P. 2006, in Society of
  Photo-Optical Instrumentation Engineers (SPIE) Conference Series, Vol. 6269,
  Society of Photo-Optical Instrumentation Engineers (SPIE) Conference Series,
  ed. I.~S. {McLean} \& M.~{Iye}, 626931

\bibitem[{{Crane} {et~al.}(2010){Crane}, {Shectman}, {Butler}, {Thompson},
  {Birk}, {Jones}, \& {Burley}}]{2010SPIE.7735E..53C}
{Crane}, J.~D., {Shectman}, S.~A., {Butler}, R.~P., {et~al.} 2010, in Society
  of Photo-Optical Instrumentation Engineers (SPIE) Conference Series, Vol.
  7735, \procspie, 773553

\bibitem[{{Crane} {et~al.}(2008){Crane}, {Shectman}, {Butler}, {Thompson}, \&
  {Burley}}]{2008SPIE.7014E..79C}
{Crane}, J.~D., {Shectman}, S.~A., {Butler}, R.~P., {Thompson}, I.~B., \&
  {Burley}, G.~S. 2008, in Society of Photo-Optical Instrumentation Engineers
  (SPIE) Conference Series, Vol. 7014, Ground-based and Airborne
  Instrumentation for Astronomy II, ed. I.~S. {McLean} \& M.~M. {Casali},
  701479

\bibitem[{{Dawson} \& {Johnson}(2018)}]{2018ARA&A..56..175D}
{Dawson}, R.~I., \& {Johnson}, J.~A. 2018, \araa, 56, 175

\bibitem[{{Donati} {et~al.}(1997){Donati}, {Semel}, {Carter}, {Rees}, \&
  {Collier Cameron}}]{1997MNRAS.291..658D}
{Donati}, J.-F., {Semel}, M., {Carter}, B.~D., {Rees}, D.~E., \& {Collier
  Cameron}, A. 1997, \mnras, 291, 658

\bibitem[{{Dotter}(2016)}]{2016ApJS..222....8D}
{Dotter}, A. 2016, \apjs, 222, 8

\bibitem[{{Eastman} {et~al.}(2019){Eastman}, {Rodriguez}, {Agol}, {Stassun},
  {Beatty}, {Vanderburg}, {Gaudi}, {Collins}, \& {Luger}}]{2019arXiv190709480E}
{Eastman}, J.~D., {Rodriguez}, J.~E., {Agol}, E., {et~al.} 2019, arXiv
  e-prints, arXiv:1907.09480

\bibitem[{{Foreman-Mackey} {et~al.}(2013){Foreman-Mackey}, {Hogg}, {Lang}, \&
  {Goodman}}]{2013PASP..125..306F}
{Foreman-Mackey}, D., {Hogg}, D.~W., {Lang}, D., \& {Goodman}, J. 2013, \pasp,
  125, 306

\bibitem[{{Gaia Collaboration} {et~al.}(2018){Gaia Collaboration}, {Brown},
  {Vallenari}, {Prusti}, {de Bruijne}, {Babusiaux}, {Bailer-Jones}, {Biermann},
  {Evans}, {Eyer}, \& et~al.}]{2018A&A...616A...1G}
{Gaia Collaboration}, {Brown}, A.~G.~A., {Vallenari}, A., {et~al.} 2018, \aap,
  616, A1

\bibitem[{{Hirano} {et~al.}(2011){Hirano}, {Suto}, {Winn}, {Taruya}, {Narita},
  {Albrecht}, \& {Sato}}]{2011ApJ...742...69H}
{Hirano}, T., {Suto}, Y., {Winn}, J.~N., {et~al.} 2011, \apj, 742, 69

\bibitem[{{Hirano} {et~al.}(2020){Hirano}, {Krishnamurthy}, {Gaidos},
  {Flewelling}, {Mann}, {Narita}, {Plavchan}, {Kotani}, {Tamura}, {Harakawa},
  {Hodapp}, {Ishizuka}, {Jacobson}, {Konishi}, {Kudo}, {Kurokawa}, {Kuzuhara},
  {Nishikawa}, {Omiya}, {Serizawa}, {Ueda}, \& {Vievard}}]{2020ApJ...899L..13H}
{Hirano}, T., {Krishnamurthy}, V., {Gaidos}, E., {et~al.} 2020, \apjl, 899, L13

\bibitem[{{Hjorth} {et~al.}(2021){Hjorth}, {Albrecht}, {Hirano}, {Winn},
  {Dawson}, {Zanazzi}, {Knudstrup}, \& {Sato}}]{2021arXiv210207677H}
{Hjorth}, M., {Albrecht}, S., {Hirano}, T., {et~al.} 2021, arXiv e-prints,
  arXiv:2102.07677

\bibitem[{{Huang} {et~al.}(2017){Huang}, {Petrovich}, \&
  {Deibert}}]{2017AJ....153..210H}
{Huang}, C.~X., {Petrovich}, C., \& {Deibert}, E. 2017, \aj, 153, 210

\bibitem[{{Huang} {et~al.}(2020){Huang}, {Vanderburg}, {P{\'a}l}, {Sha}, {Yu},
  {Fong}, {Fausnaugh}, {Shporer}, {Guerrero}, {Vanderspek}, \&
  {Ricker}}]{2020RNAAS...4..206H}
{Huang}, C.~X., {Vanderburg}, A., {P{\'a}l}, A., {et~al.} 2020, Research Notes
  of the American Astronomical Society, 4, 206

\bibitem[{{Huber} {et~al.}(2013){Huber}, {Carter}, {Barbieri}, {Miglio},
  {Deck}, {Fabrycky}, {Montet}, {Buchhave}, {Chaplin}, {Hekker},
  {Montalb{\'a}n}, {Sanchis-Ojeda}, {Basu}, {Bedding}, {Campante},
  {Christensen-Dalsgaard}, {Elsworth}, {Stello}, {Arentoft}, {Ford},
  {Gilliland}, {Handberg}, {Howard}, {Isaacson}, {Johnson}, {Karoff},
  {Kawaler}, {Kjeldsen}, {Latham}, {Lund}, {Lundkvist}, {Marcy}, {Metcalfe},
  {Silva Aguirre}, \& {Winn}}]{2013Sci...342..331H}
{Huber}, D., {Carter}, J.~A., {Barbieri}, M., {et~al.} 2013, Science, 342, 331

\bibitem[{{Jenkins} {et~al.}(2016){Jenkins}, {Twicken}, {McCauliff},
  {Campbell}, {Sanderfer}, {Lung}, {Mansouri-Samani}, {Girouard}, {Tenenbaum},
  {Klaus}, {Smith}, {Caldwell}, {Chacon}, {Henze}, {Heiges}, {Latham},
  {Morgan}, {Swade}, {Rinehart}, \& {Vanderspek}}]{2016SPIE.9913E..3EJ}
{Jenkins}, J.~M., {Twicken}, J.~D., {McCauliff}, S., {et~al.} 2016, in
  \procspie, Vol. 9913, Software and Cyberinfrastructure for Astronomy IV,
  99133E

\bibitem[{{Kamiaka} {et~al.}(2019){Kamiaka}, {Benomar}, {Suto}, {Dai},
  {Masuda}, \& {Winn}}]{2019AJ....157..137K}
{Kamiaka}, S., {Benomar}, O., {Suto}, Y., {et~al.} 2019, \aj, 157, 137

\bibitem[{{Kraus} {et~al.}(2020){Kraus}, {Le Bouquin}, {Kreplin}, {Davies},
  {Hone}, {Monnier}, {Gardner}, {Kennedy}, \& {Hinkley}}]{2020ApJ...897L...8K}
{Kraus}, S., {Le Bouquin}, J.-B., {Kreplin}, A., {et~al.} 2020, \apjl, 897, L8

\bibitem[{{Kreidberg}(2015)}]{2015PASP..127.1161K}
{Kreidberg}, L. 2015, \pasp, 127, 1161

\bibitem[{{Latham} {et~al.}(2011){Latham}, {Rowe}, {Quinn}, {Batalha},
  {Borucki}, {Brown}, {Bryson}, {Buchhave}, {Caldwell}, {Carter},
  {Christiansen}, {Ciardi}, {Cochran}, {Dunham}, {Fabrycky}, {Ford}, {Gautier},
  {Gilliland}, {Holman}, {Howell}, {Ibrahim}, {Isaacson}, {Jenkins}, {Koch},
  {Lissauer}, {Marcy}, {Quintana}, {Ragozzine}, {Sasselov}, {Shporer},
  {Steffen}, {Welsh}, \& {Wohler}}]{2011ApJ...732L..24L}
{Latham}, D.~W., {Rowe}, J.~F., {Quinn}, S.~N., {et~al.} 2011, \apjl, 732, L24

\bibitem[{{Mandel} \& {Agol}(2002)}]{2002ApJ...580L.171M}
{Mandel}, K., \& {Agol}, E. 2002, \apjl, 580, L171

\bibitem[{{Mann} {et~al.}(2019){Mann}, {Dupuy}, {Kraus}, {Gaidos}, {Ansdell},
  {Ireland}, {Rizzuto}, {Hung}, {Dittmann}, {Factor}, {Feiden}, {Martinez},
  {Ru{\'\i}z-Rodr{\'\i}guez}, \& {Thao}}]{2019ApJ...871...63M}
{Mann}, A.~W., {Dupuy}, T., {Kraus}, A.~L., {et~al.} 2019, \apj, 871, 63

\bibitem[{{Mann} {et~al.}(2020){Mann}, {Johnson}, {Vanderburg}, {Kraus},
  {Rizzuto}, {Wood}, {Bush}, {Rockcliffe}, {Newton}, {Latham}, {Mamajek},
  {Zhou}, {Quinn}, {Thao}, {Benatti}, {Cosentino}, {Desidera}, {Harutyunyan},
  {Lovis}, {Mortier}, {Pepe}, {Poretti}, {Wilson}, {Kristiansen}, {Gagliano},
  {Jacobs}, {LaCourse}, {Omohundro}, {Schwengeler}, {Terentev}, {Kane}, {Hill},
  {Rabus}, {Esquerdo}, {Berlind}, {Collins}, {Murawski}, {Sallam}, {Aitken},
  {Massey}, {Ricker}, {Vanderspek}, {Seager}, {Winn}, {Jenkins}, {Barclay},
  {Caldwell}, {Dragomir}, {Doty}, {Glidden}, {Tenenbaum}, {Torres}, {Twicken},
  \& {Villanueva}}]{2020AJ....160..179M}
{Mann}, A.~W., {Johnson}, M.~C., {Vanderburg}, A., {et~al.} 2020, \aj, 160, 179

\bibitem[{{Marcy} \& {Butler}(1992)}]{1992PASP..104..270M}
{Marcy}, G.~W., \& {Butler}, R.~P. 1992, \pasp, 104, 270

\bibitem[{{Martioli} {et~al.}(2020){Martioli}, {H{\'e}brard}, {Moutou},
  {Donati}, {Artigau}, {Cale}, {Cook}, {Dalal}, {Delfosse}, {Forveille},
  {Gaidos}, {Plavchan}, {Berberian}, {Carmona}, {Cloutier}, {Doyon},
  {Fouqu{\'e}}, {Klein}, {Lecavelier des Etangs}, {Manset}, {Morin}, {Tanner},
  {Teske}, \& {Wang}}]{2020A&A...641L...1M}
{Martioli}, E., {H{\'e}brard}, G., {Moutou}, C., {et~al.} 2020, \aap, 641, L1

\bibitem[{{Masuda} \& {Winn}(2020)}]{2020AJ....159...81M}
{Masuda}, K., \& {Winn}, J.~N. 2020, \aj, 159, 81

\bibitem[{{McLaughlin}(1924)}]{1924ApJ....60...22M}
{McLaughlin}, D.~B. 1924, \apj, 60, 22

\bibitem[{{Montet} {et~al.}(2020){Montet}, {Feinstein}, {Luger}, {Bedell},
  {Gully-Santiago}, {Teske}, {Wang}, {Butler}, {Flowers}, {Shectman}, {Crane},
  \& {Thompson}}]{2020AJ....159..112M}
{Montet}, B.~T., {Feinstein}, A.~D., {Luger}, R., {et~al.} 2020, \aj, 159, 112

\bibitem[{{Oshagh} {et~al.}(2013){Oshagh}, {Boisse}, {Bou{\'e}}, {Montalto},
  {Santos}, {Bonfils}, \& {Haghighipour}}]{2013A&A...549A..35O}
{Oshagh}, M., {Boisse}, I., {Bou{\'e}}, G., {et~al.} 2013, \aap, 549, A35

\bibitem[{{Oshagh} {et~al.}(2018){Oshagh}, {Triaud}, {Burdanov}, {Figueira},
  {Reiners}, {Santos}, {Faria}, {Boue}, {D{\'\i}az}, {Dreizler}, {Boldt},
  {Delrez}, {Ducrot}, {Gillon}, {Guzman Mesa}, {Jehin}, {Khalafinejad}, {Kohl},
  {Serrano}, \& {Udry}}]{2018A&A...619A.150O}
{Oshagh}, M., {Triaud}, A.~H.~M.~J., {Burdanov}, A., {et~al.} 2018, \aap, 619,
  A150

\bibitem[{{Palle} {et~al.}(2020){Palle}, {Oshagh}, {Casasayas-Barris},
  {Hirano}, {Stangret}, {Luque}, {Strachan}, {Gaidos}, {Anglada-Escude},
  {Plavchan}, \& {Addison}}]{2020A&A...643A..25P}
{Palle}, E., {Oshagh}, M., {Casasayas-Barris}, N., {et~al.} 2020, \aap, 643,
  A25

\bibitem[{{Perryman} {et~al.}(1997){Perryman}, {Lindegren}, {Kovalevsky},
  {Hog}, {Bastian}, {Bernacca}, {Creze}, {Donati}, {Grenon}, {Grewing}, {van
  Leeuwen}, {van der Marel}, {Mignard}, {Murray}, {Le Poole}, {Schrijver},
  {Turon}, {Arenou}, {Froeschle}, \& {Petersen}}]{1997AA...323L..49P}
{Perryman}, M.~A.~C., {Lindegren}, L., {Kovalevsky}, J., {et~al.} 1997, \aap,
  500, 501

\bibitem[{{Roettenbacher} {et~al.}(2017){Roettenbacher}, {Monnier}, {Korhonen},
  {Harmon}, {Baron}, {Hackman}, {Henry}, {Schaefer}, {Strassmeier}, {Weber}, \&
  {ten Brummelaar}}]{2017ApJ...849..120R}
{Roettenbacher}, R.~M., {Monnier}, J.~D., {Korhonen}, H., {et~al.} 2017, \apj,
  849, 120

\bibitem[{{Rossiter}(1924)}]{1924ApJ....60...15R}
{Rossiter}, R.~A. 1924, \apj, 60, 15

\bibitem[{{Rubenzahl} {et~al.}(2021){Rubenzahl}, {Dai}, {Howard}, {Chontos},
  {Giacalone}, {Lubin}, {Rosenthal}, {Isaacson}, {Batalha}, {Crossfield},
  {Dressing}, {Fulton}, {Huber}, {Kane}, {Petigura}, {Robertson}, {Roy},
  {Weiss}, {Beard}, {Hill}, {Mayo}, {Mocnik}, {Murphy}, \&
  {Scarsdale}}]{2021AJ....161..119R}
{Rubenzahl}, R.~A., {Dai}, F., {Howard}, A.~W., {et~al.} 2021, \aj, 161, 119

\bibitem[{{Sanchis-Ojeda} \& {Winn}(2011)}]{2011ApJ...743...61S}
{Sanchis-Ojeda}, R., \& {Winn}, J.~N. 2011, \apj, 743, 61

\bibitem[{{Skrutskie} {et~al.}(2006){Skrutskie}, {Cutri}, {Stiening},
  {Weinberg}, {Schneider}, {Carpenter}, {Beichman}, {Capps}, {Chester},
  {Elias}, {Huchra}, {Liebert}, {Lonsdale}, {Monet}, {Price}, {Seitzer},
  {Jarrett}, {Kirkpatrick}, {Gizis}, {Howard}, {Evans}, {Fowler}, {Fullmer},
  {Hurt}, {Light}, {Kopan}, {Marsh}, {McCallon}, {Tam}, {Van Dyk}, \&
  {Wheelock}}]{2006AJ....131.1163S}
{Skrutskie}, M.~F., {Cutri}, R.~M., {Stiening}, R., {et~al.} 2006, \aj, 131,
  1163

\bibitem[{{Stefansson} {et~al.}(2020){Stefansson}, {Mahadevan}, {Maney},
  {Ninan}, {Robertson}, {Rajagopal}, {Haase}, {Allen}, {Ford}, {Winn},
  {Wolfgang}, {Dawson}, {Wisniewski}, {Bender}, {Ca{\~n}as}, {Cochran},
  {Diddams}, {Fredrick}, {Halverson}, {Hearty}, {Hebb}, {Kanodia}, {Levi},
  {Metcalf}, {Monson}, {Ramsey}, {Roy}, {Schwab}, {Terrien}, \&
  {Wright}}]{2020AJ....160..192S}
{Stefansson}, G., {Mahadevan}, S., {Maney}, M., {et~al.} 2020, \aj, 160, 192

\bibitem[{{Triaud}(2018)}]{2018haex.bookE...2T}
{Triaud}, A. H.~M.~J. 2018, {The Rossiter-McLaughlin Effect in Exoplanet
  Research}, 2

\bibitem[{{Yee} {et~al.}(2018){Yee}, {Petigura}, {Fulton}, {Knutson},
  {Batygin}, {Bakos}, {Hartman}, {Hirsch}, {Howard}, {Isaacson}, {Kosiarek},
  {Sinukoff}, \& {Weiss}}]{2018AJ....155..255Y}
{Yee}, S.~W., {Petigura}, E.~A., {Fulton}, B.~J., {et~al.} 2018, \aj, 155, 255

\bibitem[{{Zhou} {et~al.}(2020){Zhou}, {Winn}, {Newton}, {Quinn}, {Rodriguez},
  {Mann}, {Rizzuto}, {Vanderburg}, {Huang}, {Latham}, {Teske}, {Wang},
  {Shectman}, {Butler}, {Crane}, {Thompson}, {Henry}, {Paredes}, {Jao},
  {James}, \& {Hinojosa}}]{2020ApJ...892L..21Z}
{Zhou}, G., {Winn}, J.~N., {Newton}, E.~R., {et~al.} 2020, \apjl, 892, L21

\bibitem[{{Zhou} {et~al.}(2021){Zhou}, {Quinn}, {Irwin}, {Huang}, {Collins},
  {Bouma}, {Khan}, {Landrigan}, {Vanderburg}, {Rodriguez}, {Latham}, {Torres},
  {Douglas}, {Bieryla}, {Esquerdo}, {Berlind}, {Calkins}, {Buchhave},
  {Charbonneau}, {Collins}, {Kielkopf}, {Jensen}, {Tan}, {Hart}, {Carter},
  {Stockdale}, {Ziegler}, {Law}, {Mann}, {Howell}, {Matson}, {Scott}, {Furlan},
  {White}, {Hellier}, {Anderson}, {West}, {Ricker}, {Vanderspek}, {Seager},
  {Jenkins}, {Winn}, {Mireles}, {Rowden}, {Yahalomi}, {Wohler}, {Brasseur},
  {Daylan}, \& {Col{\'o}n}}]{2021AJ....161....2Z}
{Zhou}, G., {Quinn}, S.~N., {Irwin}, J., {et~al.} 2021, \aj, 161, 2

\end{thebibliography}

\end{document}